\documentclass[11pt,twoside]{article}
\usepackage{asp2010}

\resetcounters

\begin{document}

\title{Monitoring the Polarimetric Variability of $\delta$ Scorpii}
\author{D.~Bednarski and A.~C.~Carciofi
\affil{Instituto de Astronomia, Geoci\^{e}ncias e Ci\^{e}ncias Atmosf\'{e}ricas,\\
Universidade de S\~{a}o Paulo, Rua do Mat\~{a}o 1226, 05508-900,\\ 
Cidade Universit\'{a}ria, S\~{a}o Paulo, SP, BRAZIL}}

\begin{abstract}
The Be star $\delta$~Scorpii is an interesting binary system, whose primary companion created a circumstellar disk after the periastron passage of the secondary in 2000, being since then classified as Be. This work presents the results of a long-term monitoring of this star in broad-band imaging polarimetry.
The observational data collected since 2006 in the Pico dos Dias Observatory (Brazil) show a variable polarization that seems to correlate with the photometric light curve. From this data we see that the disk density varied since 2006; furthermore, the data suggests that there was some disturbance of the disk during the last periastron passage in July, 2011. 
\end{abstract}


\section{Introduction}
            
Delta Scorpii is a binary system, whose primary component is a Be star and the secondary's nature is still unknown (see Miroshnichenko, these proceedings). 
After the 1990 passage of the secondary companion at the periastron there was some detectable line emission (suggestive of circumstellar material), which became more intense in next passage, in 2000. Since then, the primary star has developed a quite large and strong disk \citep[e.g.,][]{carciofi_2006}.


Be stars have an intrinsic linear polarization, which arises from the scattering of starlight from the free electrons in disk. Polarization has  since long been used in Be star research as a disk diagnostics \citep[see, e.g.,][]{Wood_1997}; recently, \citet{Draper_2011} have studied the usefulness of polarization as a diagnostic of disk evolution. 

In this contribution, we show the results of a polarization monitoring of $\delta$ Sco carried out since 2006.

\section{Observations}
  
The polarimetric data were obtained primarily in the 0.60\,m Boller \& Chivens telescope at the Pico dos Dias Observatory (Minas Gerais, Brazil). The Perkin-Elmer (1.60\,m) and Zeiss (0.60\,m) telescopes were used in few missions.

The instrument used was the IAGPOL polarimeter, capable of obtaining high precision imaging polarimetry in the UBVRI filters. \citet{Carciofi_2007} have demonstrated that IAGPOL is photon-noise limited up to polarization levels of $\sim 0.005\%$. A typical uncertainly in our measurements is $\sim 0.015\%$, but this varies due to a number of reasons, the main one is the different observing conditions (cloud cover). So far not all available data in the 2006 -- 2009 period has been reduced yet, so this contribution presents only a sub-set of our observations in this period.

The data reduction was performed within the IRAF environment. For details on the reduction and observing strategy see \citet{Carciofi_2007} and references therein.

\subsection{Interstellar Polarization}

Usually, a very tricky part of polarization studies is to obtain an estimate of the interstellar polarization (ISP) along the line of sight. For $\delta$ Sco we could not find a reliable estimate of the ISP in the literature, so we measured the polarization of few main-sequence field stars in order to determine this quantity. The analysis is still ongoing, but our preliminary results indicate that the field star HD~142705 (spectral type A0V) is a good proxy for the ISP. The BVRI polarization of this star is shown in Table~1.


\begin{table}[!h]
\label{tab:hd}
\caption{Polarization values of the field star HD~142705, used here as a proxy of the interstellar polarization.}
\smallskip
\begin{center}
{\small
\begin{tabular}{ccc}
\tableline
\noalign{\smallskip}
 Filter & P (\%) & $\theta$ (deg) \\
\noalign{\smallskip}
\tableline
\noalign{\smallskip}
B & $0.138 \pm 0.011$ & $116.0 \pm 2.3$ \\
V & $0.159 \pm 0.016$ & $120.8 \pm 2.9$ \\
R & $0.194 \pm 0.020$ & $122.1 \pm 2.9$ \\
I & $0.143 \pm 0.016$ & $123.2 \pm 3.3$ \\
\noalign{\smallskip}
\tableline
\end{tabular}
}
\end{center}
\end{table}

\begin{figure}[!b]
\begin{center}
  \includegraphics[scale=0.57]{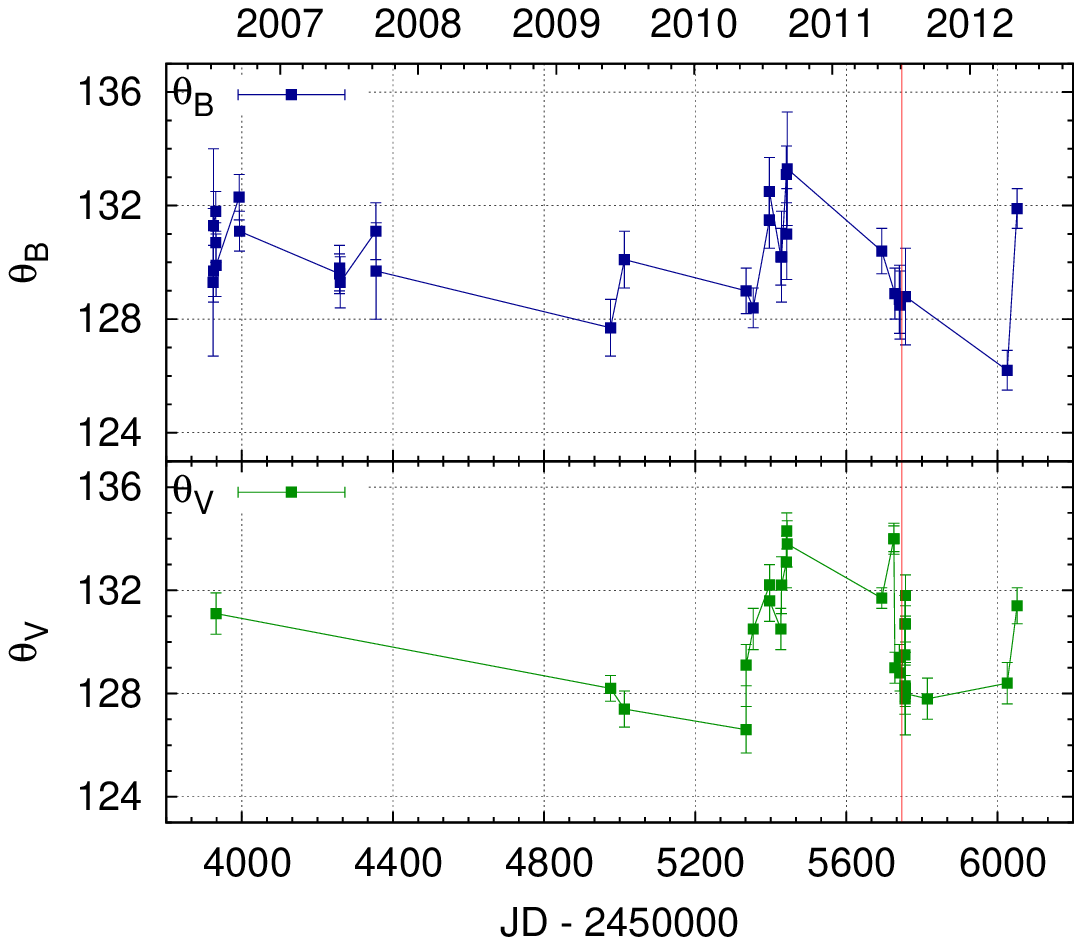}
  \hspace{10px}
  \includegraphics[scale=0.57]{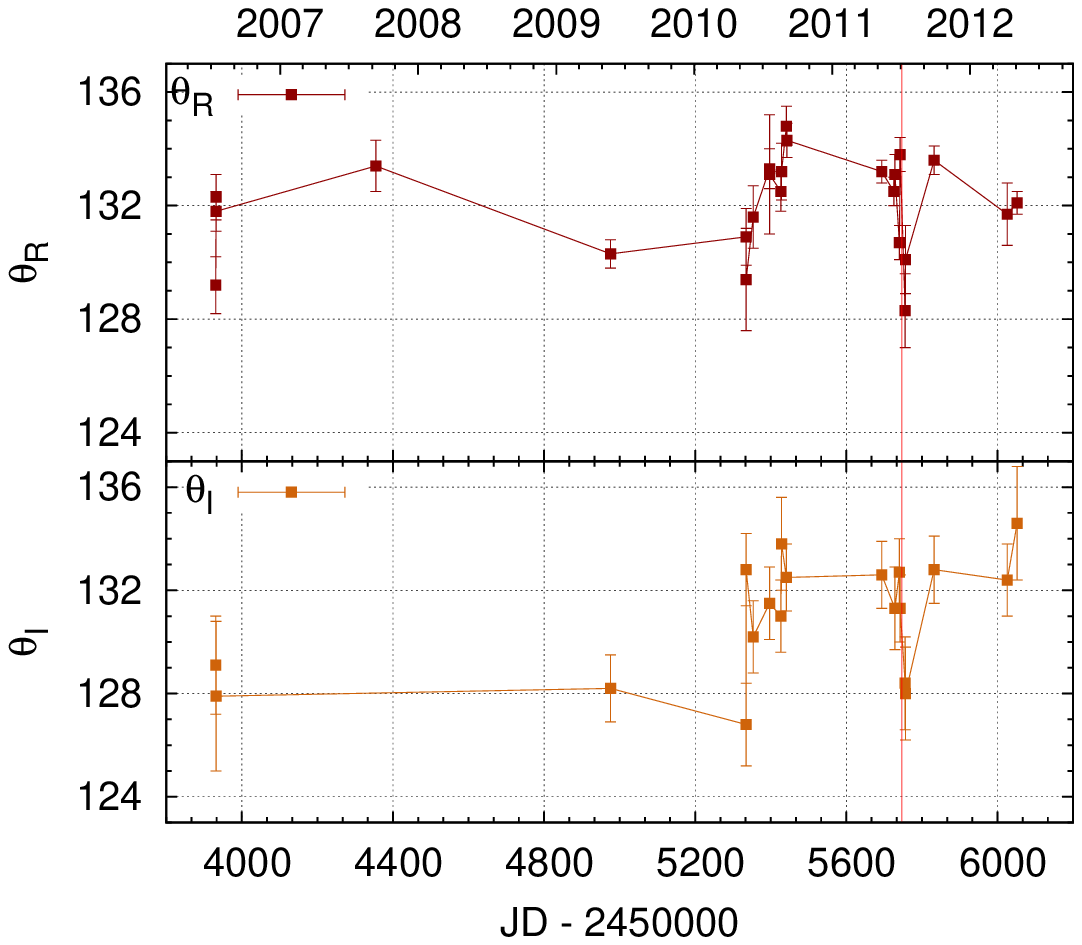}
  \caption{Intrinsic polarization angle of $\delta$ Sco in the BVRI filters. The vertical red line marks the epoch of the periastron passage.}
  \label{fig:theta}
\end{center}
\end{figure}

\begin{figure}[!h]
\begin{center}
  \includegraphics[scale=0.59]{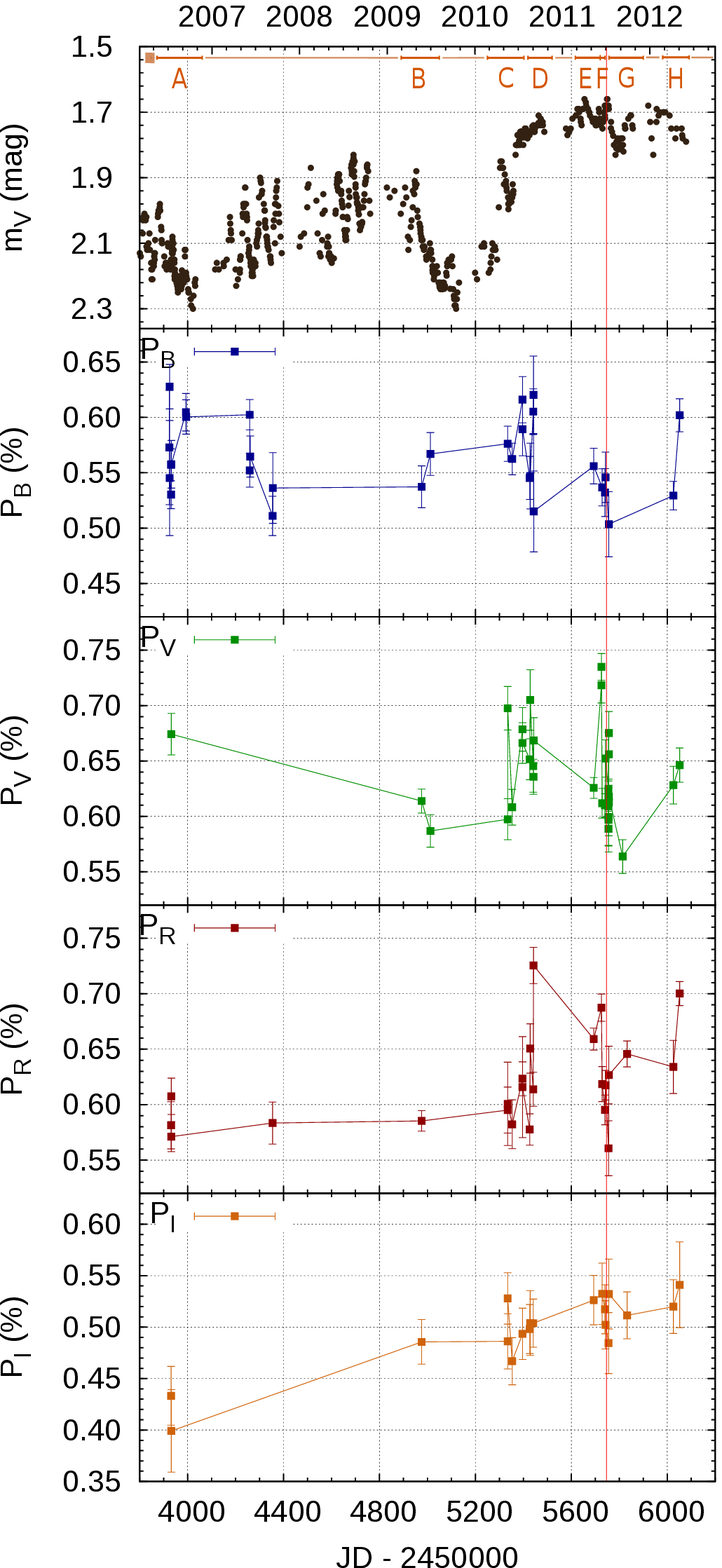}
  \hspace{10px}
  \includegraphics[scale=0.59]{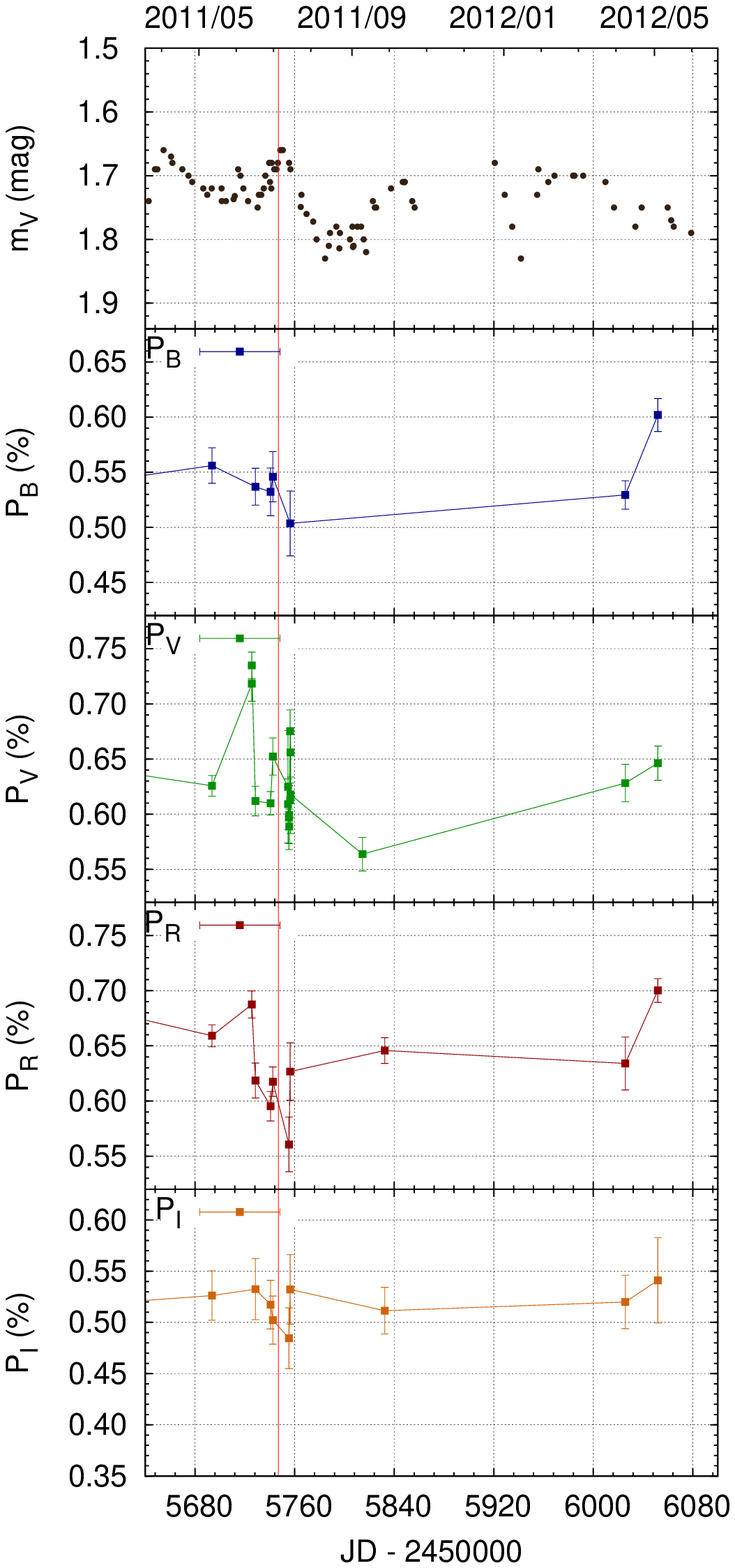}
  \caption{Top to bottom, on both sides: Light curve of $\delta$ Scorpii (S. Otero, priv. comm.) and intrinsic linear polarization in the B, V, R and I filters. The vertical red line marks the epoch of the periastron passage.}
  \label{fig:pol1}
\end{center}
\end{figure}

\begin{figure}[!ht]
\begin{center}
  \includegraphics[scale=0.56]{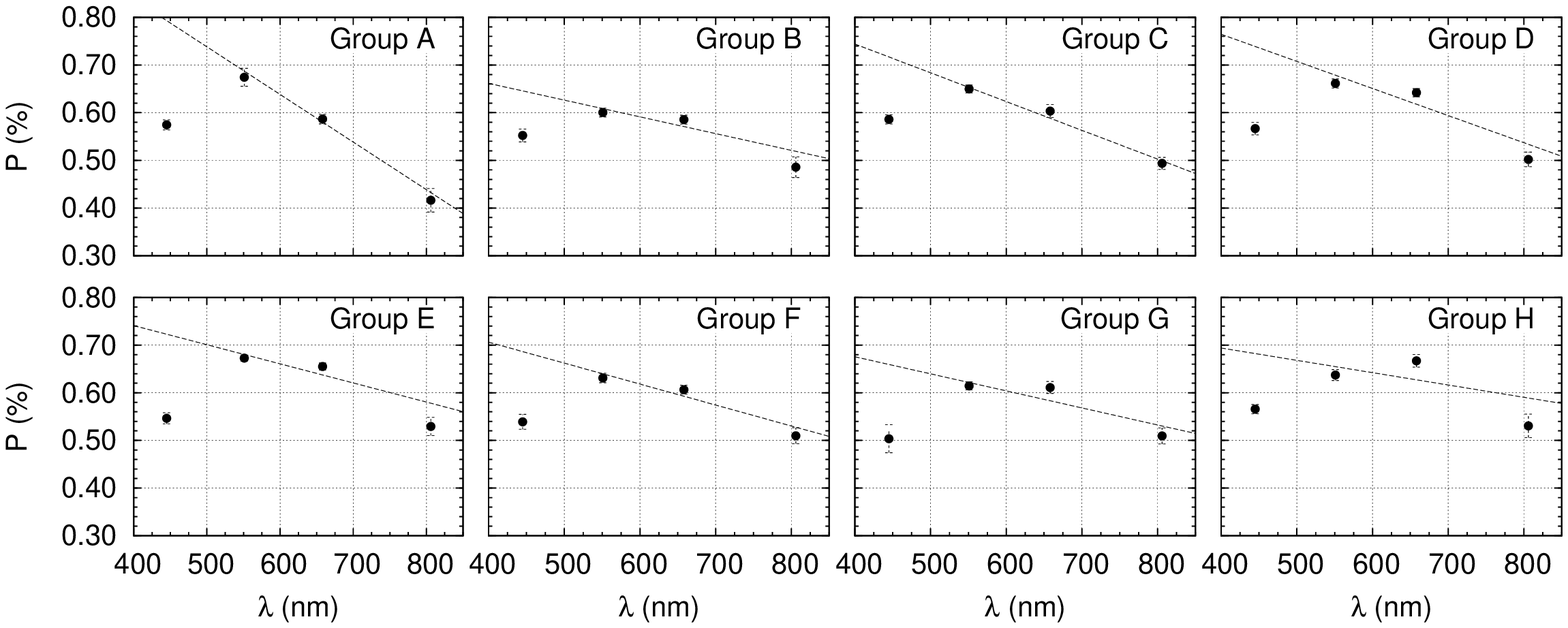}
  \caption{Polarized spectrum of $\delta$ Scorpii at different epochs, marked in Fig.~\ref{fig:pol1}.}
  \label{fig:spec}
\end{center}
\end{figure}

\section{Results and Conclusions}

The preliminary results of our polarization monitoring are presented below. In all plots the results have been corrected for the ISP, but it is important to recall that this correction is still preliminary. 

In Fig.~\ref{fig:theta} we show the polarization angle (PA) for the BVRI filters. It is hard from the data to judge about possible variability, but one can safely put an upper limit of $\sim 2\deg$ to variations of the PA. Noteworthy is the absence of (large) PA variation during the periastron passage. Models of the tidal interaction between the disk and the secondary predict important disk warping during the periastron passage, depending on the orbital parameters and the orientation of the disk with respect to the orbital plane (see Okazaki, these proceedings). A disk warping would result in variations of the PA; the fact that none (or little) variation was observed gives important constraints to the model.

The BVRI polarization curves are shown in Fig.~\ref{fig:pol1}, along with the light curve of $\delta$ Sco. These photometric data were obtained by S. Otero (visual photometry), B. Fraser and D. West (PEP photometry), and T. Moon (CCD photometry). The average error is about 0.05 mag.  One of the main findings of this work is that at times the polarization seems to be correlated with the photometry, whereas at others it is anti-correlated. This behavior can be understood if one considers the fact that those observables trace the disk density at different disk locations: visual photometry is sensitive to the inner disk structure while polarization tracks the structure of the disk over larger areas \citep[see][]{Carciofi_2011}.  We expect, therefore, that the combination of photometry and polarization will impose quite stringent constraints on any model that try to explain the disk evolution.

The observed trends are indicative of a complex evolution of disk surface density over time. This is further illustrated by the temporal changes in the slope of the polarized spectrum, shown in Fig.~\ref{fig:spec}   for eight different epochs (marked in Fig.~\ref{fig:pol1}). Least squares fits were applied on the VRI data,  aiming at illustrating the temporal evolution of the polarization slope.
This variation suggests that the density of disk is variable.

Finally, there are no signs of total destruction of the disk during the recent periastron passage. The polarization decreased slightly near the event, but quickly returned to its pre-periastron value.

\acknowledgements D.B. acknowledges support from CNPq (grant 146314/2011-5).  A.C.C. acknowledges support from CNPq (grant 308985/2009-5) and Fapesp (grant 2010/19029-0). D.B. thanks IAG and the organizing committee for the financial aid. 

\bibliography{p03_bednarski}

\begin{thebibliography}{}
\expandafter\ifx\csname natexlab\endcsname\relax\def\natexlab#1{#1}\fi
\expandafter\ifx\csname url\endcsname\relax
  \def\url#1{\texttt{#1}}\fi
\expandafter\ifx\csname urlprefix\endcsname\relax\def\urlprefix{URL }\fi
\providecommand{\eprint}[2][]{\url{#2}}

\bibitem[{{Carciofi}(2011)}]{Carciofi_2011}
{Carciofi}, A.~C. 2011, in IAU Symposium, edited by C.~{Neiner}, G.~{Wade},
  G.~{Meynet}, \& G.~{Peters}, vol. 272 of IAU Symposium, 325.
  \eprint{1009.3969}

\bibitem[{{Carciofi} et~al.(2007){Carciofi}, {Magalh{\~a}es}, {Leister},
  {Bjorkman}, \& {Levenhagen}}]{Carciofi_2007}
{Carciofi}, A.~C., {Magalh{\~a}es}, A.~M., {Leister}, N.~V., {Bjorkman}, J.~E.,
  \& {Levenhagen}, R.~S. 2007, \apjl, 671, L49

\bibitem[{{Carciofi} et~al.(2006){Carciofi}, {Miroshnichenko}, {Kusakin},
  {Bjorkman}, {Bjorkman}, {Marang}, {Kuratov}, {Garc{\'{\i}}a-Lario},
  {Calder{\'o}n}, {Fabregat}, \& {Magalh{\~a}es}}]{carciofi_2006}
{Carciofi}, A.~C., {Miroshnichenko}, A.~S., {Kusakin}, A.~V., {Bjorkman},
  J.~E., {Bjorkman}, K.~S., {Marang}, F., {Kuratov}, K.~S.,
  {Garc{\'{\i}}a-Lario}, P., {Calder{\'o}n}, J.~V.~P., {Fabregat}, J., \&
  {Magalh{\~a}es}, A.~M. 2006, \apj, 652, 1617

\bibitem[{{Draper} et~al.(2011){Draper}, {Wisniewski}, {Bjorkman}, {Haubois},
  {Carciofi}, {Bjorkman}, {Meade}, \& {Okazaki}}]{Draper_2011}
{Draper}, Z.~H., {Wisniewski}, J.~P., {Bjorkman}, K.~S., {Haubois}, X.,
  {Carciofi}, A.~C., {Bjorkman}, J.~E., {Meade}, M.~R., \& {Okazaki}, A. 2011,
  \apjl, 728, L40

\bibitem[{{Wood} et~al.(1997){Wood}, {Bjorkman}, \& {Bjorkman}}]{Wood_1997}
{Wood}, K., {Bjorkman}, K.~S., \& {Bjorkman}, J.~E. 1997, \apj, 477, 926

\end{thebibliography}

\end{document}